\documentclass[11pt]{article}

\usepackage[final]{acl}
\usepackage{times}
\usepackage{latexsym}
\usepackage[T1]{fontenc}
\usepackage[utf8]{inputenc}
\usepackage{microtype}
\usepackage{inconsolata}
\usepackage{graphicx}
\usepackage{fontawesome}
\usepackage{booktabs}
\usepackage{bm}
\usepackage{xcolor}
\usepackage{listings}

\definecolor{manniBg}{HTML}{F0F3F3} 
\definecolor{manniString}{HTML}{008080}
\definecolor{manniKey}{HTML}{000080}  
\definecolor{manniNumber}{HTML}{000000}
\definecolor{manniPunct}{HTML}{555555}

\lstdefinelanguage{json}{
    basicstyle=\small\ttfamily,
    backgroundcolor=\color{manniBg},
    frame=single,
    rulecolor=\color{gray!30}, 
    framesep=5pt,
    breaklines=true,
    showstringspaces=false,
    upquote=true,
    morestring=[b]", 
    stringstyle=\color{manniString},
    morekeywords={true,false,null},
    keywordstyle=\color{manniKey}\bfseries,
    literate=
     *{:}{{{\color{manniPunct}{:}}}}{1}
      {,}{{{\color{manniPunct}{,}}}}{1}
      {\{}{{{\color{manniPunct}{\{}}}}{1}
      {\}}{{{\color{manniPunct}{\}}}}}{1}
      {[}{{{\color{manniPunct}{[}}}}{1}
      {]}{{{\color{manniPunct}{]}}}}{1},
}

\DeclareCaptionType{listing}[Listing][List of Listings]

\newcommand{\amc}{\texttt{amc}}

\title{\amc: The Automated Mission Classifier for Telescope Bibliographies}

\author{John F. Wu$^*$ \hspace{1em}
  Joshua E. G. Peek \hspace{1em}
  Sophie J. Miller  \\
 \textbf{
  Jenny Novacescu  \hspace{1em}
  Achu J. Usha \hspace{1em}
  Christopher A. Wilkinson} \\
  Space Telescope Science Institute \\
  3700 San Martin Dr \\
  Baltimore, MD 21218 USA \\ 
  $^*$\texttt{jfwu@stsci.edu}
}

\begin{document}
\maketitle

\begin{abstract}
Telescope bibliographies record the pulse of astronomy research by capturing publication statistics and citation metrics for telescope facilities. 
Robust and scalable bibliographies ensure that we can measure the scientific impact of our facilities and archives.
However, the growing rate of publications threatens to outpace our ability to manually label astronomical literature.
We therefore present the Automated Mission Classifier (\amc{}), a tool that uses large language models (LLMs) to identify and categorize telescope references by processing large quantities of paper text. 
A modified version of \amc{} performs well on the TRACS Kaggle challenge, achieving a macro $F_1$ score of $0.84$ on the held-out test set. 
\amc{} is valuable for other telescopes beyond TRACS; we developed the initial software for identifying papers that featured scientific results by NASA missions.
Additionally, we investigate how \amc{} can also be used to interrogate historical datasets and surface potential label errors.
Our work demonstrates that LLM-based applications offer powerful and scalable assistance for library sciences.
\href{https://github.com/jwuphysics/automated-mission-classifier}{\faGithub}
\end{abstract}

\section{Introduction}

Telescope bibliographies provide one way to measure the scientific productivity of our astronomical facilities.
Bibliometrics can quantify how often telescopes are discussed in scientific publications, e.g., through passing mentions or via detailed scientific analyses that originate from data taken by each telescope.
Although these quantitative analyses are vital for assessing the impact of our scientific investments, they hinge on complete, homogeneous bibliographies, which can be expensive and onerous to manually curate.
Librarians, archive scientists, and bibliographers maintain telescope bibliographies by consistently tracking publications, extracting metadata, and labeling the scientific \textit{intent} of each telescope reference for all papers \citep[see, e.g.,][]{2015ASPC..492...99L,2024OJAp....7E..85O}.
Complete observatory bibliographies enable us to investigate publication rates, and citation statistics, links between publications and observing proposals, data product usage metrics, and archival science impact \citep[e.g.,][for HST]{2010PASP..122..808A}.

There is more scientific literature than ever before \citep[notwithstanding gender-disparate impacts from the recent pandemic,][]{2023NatAs...7..105B}. Some of this increase accompanies a general rise of publication rates throughout academia \citep{hanson2024strain}. Additionally, very recent growth in publication rates may stem from the advent of large language models (LLMs), which can lower the barrier to writing papers \citep[e.g.,][]{2024sais.conf..241A}. These trends suggest that we need a sustainable solution for producing telescope bibliographies amid the deluge of astronomy papers.

LLMs can also be useful for compiling telescope bibliographies at scale: 
artificial intelligence (AI) systems are highly scalable, and are now adept at processing large amounts of text inputs.
Modern LLMs can complete many tasks \textit{without any optimization}, instead relying solely on emergent capabilities like in-context learning \citep[e.g.,][]{gpt3}.
With frontier AI labs now deploying LLMs as a service, we can easily leverage simple API (Application Programming Interface) calls and design software around cutting-edge LLMs.

Before deploying an automated bibliography system, we must first ensure that its performance is \textit{robust}.
To this end, we present and evaluate the Automated Mission Classifier (\texttt{amc}), an LLM-powered, bibliometric tool for identifying telescopes or NASA missions in the literature. We adapt \amc{} for a specific shared task, TRACS (Section~\ref{sec:tracs}); in the Appendix, we note that similar systems are already in operations for JWST (Appendix~\ref{sec:jwst-preprints}) and can be used for archival science with other telescopes (Appendix~\ref{sec:other-bib}). In Section~\ref{sec:amc}, we describe the software's system design, and we present results in Section~\ref{sec:results}. 
In Section~\ref{sec:discussion}, we discuss how observatory bibliographers can leverage AI to compute bibliometrics at scale, assess (historical) data quality, and upgrade the LLM systems.
We provide publicly available code on Github: \url{https://github.com/jwuphysics/automated-mission-classifier}.

\begin{quote}
``To LLMs! The cause of, and solution to, all of bibliographers' problems.''\footnote{{Quote adapted from \textit{The Simpsons} \citep{Simpsons:HomerVs18th}.}}
\end{quote}

\section{The TRACS Shared Task} \label{sec:tracs}

The Telescope Reference and Astronomy Categorization Shared task (TRACS) is a data challenge organized as part of the 2025 Workshop for Artificial Intelligence for Scientific Publications (WASP; \citealt{grezes-etal-2025-tracs}) at IJCNLP-AACL.\footnote{For details on The International Joint Conference on Natural Language Processing \& Asia-Pacific Chapter of the Association for Computational Linguistics (IJCNLP-AACL), see \url{https://2025.aaclnet.org/}.} 
The task consists of classifying astronomy papers into at least one of four categories: \texttt{science}, \texttt{instrumentation}, \texttt{mention}, or \texttt{not\_telescope}.

In this data challenge, papers are decomposed into several fields (including the title, abstract, and ``body'' full text) and, based on keyword filtering, labeled with a candidate telescope name (\texttt{CHANDRA}, \texttt{HST}, \texttt{JWST}, or \texttt{NONE}).
The objective is to predict the boolean labels for all paper categories for each of the provided bibcode + telescope combinations. However, it is important to note that the candidate telescope name may be mislabeled, and that certain paper categories impose constraints on the others (i.e., a single paper + telescope can have True labels for both \texttt{science} and \texttt{instrumentation}, but cannot for both \texttt{science} and \texttt{not\_telescope}).

Training and test data sets, in CSV format with 80,385 and 9,194 entries respectively, are provided for the shared task. 
To participate in the challenge, entrants must submit 9,194 test-set predictions via Kaggle\footnote{\url{https://www.kaggle.com/competitions/tracs-wasp-2025}} and have their predictions evaluated.
The test outputs are scored according to the average between the macro $F_1$ score of the telescope labels and the macro $F_1$ score of paper labels; each class is weighted equally. 
Note that \texttt{NONE} is a valid telescope class and \texttt{not\_telescope} is a valid paper class.
In the subsections below, we note some details that we considered important for our submission.

\subsection{Input Data}

The full list of columns in the train data set include: (0) ID, (1) bibcode, (2) telescope, (3) author, (4) year, (5) title, (6) abstract, (7) body, (8) acknowledgments, and (9) grants, (10) \texttt{science}, (11) \texttt{instrumentation}, (12) \texttt{mention}, and (13) \texttt{not\_telescope}.
The test dataset does not include column (2) or columns (10) through (13).
However, a preliminary telescope label \textit{is} implicitly named in column (0), as the ID is simply the concatenation of \texttt{\{bibcode\}\_\{telescope\}}.

Some rows in the datasets are missing: 3\% of the test data set is missing an abstract, 19\% does not have full-body text, and $>90$\% does not have text under the grants column. Incomplete data are likely due to a combination of parsing errors (e.g., correctly parsing out grants/acknowledgments) and publisher restrictions. Issues with publisher agreements tend to impact certain journals or publication venues (i.e., demarcated by their ``bibstem'' entries); in many of these cases, the body text is completely absent. Nonetheless, classifications can sometimes still be made on the basis of just the title and abstract (but see Appendix~\ref{sec:other-bib}).

Some of the input data may not be helpful. For example, the list of authors is unlikely to yield useful indicators of the paper classification, and may even produce false positives, as ``Webb'' or ``Chandra'' can show up as (sub-word) names of authors. Likewise, ``Hubble'' can often show up in the acknowledgments, e.g. due to funding acknowledgments from the NASA Hubble Fellowship Program. Thus, it is imperative to design a language modeling system that can flexibly understand the context surrounding telescope detections.

\subsection{Paper Types}

Establishing a common definition for paper types is a nontrivial task. When tasking human bibliographers to classify papers, e.g., identify \texttt{science} papers, disagreements often arise about the precise definition of a \texttt{science} paper. 

In order to implement a useful LLM system for automated classification, it is necessary to unambiguously define the labels. 
\citet[][]{2024OJAp....7E..85O} issue the following guidance on \texttt{science} papers:
\begin{quote}
``To qualify as a science paper, it must be apparent that data or data product(s) from the observatory were used and that the data or data product(s) formed the basis for reaching a new scientific conclusion.''
\end{quote}
The authors recognize that these definitions must be continually updated.

Indeed, the taxonomy should serve the telescope or mission. Existing schemes may not be sufficient to characterize all of the edge cases, and new categories may arise. As a concrete example, STScI established a \textit{data-influenced} category in 2019 for papers that indirectly rely on data or products, but do not directly analyze data or use data products. 
In the TRACS challenge taxonomy, data-influenced papers would generally be labeled under the \texttt{mention} category.

As part of the shared task, the TRACS website provided a narrative format description of the different paper types \citep{grezes-etal-2025-tracs}. We used an LLM (\texttt{claude-sonnet-4.1}) to process this text in order to create a user prompt that includes definitions and examples of each paper type (which is manually updated and described in more detail in Section~\ref{sec:amc}). The full prompts can be found in the Github repository, and we have copied the paper type definitions here (note that we remove the markdown text formatting for human readability):
\begin{itemize}
    \item \texttt{science}: Paper directly uses \{\{telescope\}\} data (new or published) to obtain new scientific results in this paper.
    \item \texttt{instrumentation}: Paper describes new instrument science or engineering.
    \item \texttt{mention}: Paper references the telescope but does not produce new scientific results.
    \item \texttt{not\_telescope}: Paper includes references that are false positives -- names that look like the telescope but refer to something completely different.
\end{itemize}

\section{The Automated Mission Classifier (\amc)} \label{sec:amc}

\begin{figure*}[t]
    \centering
    \includegraphics[width=\linewidth]{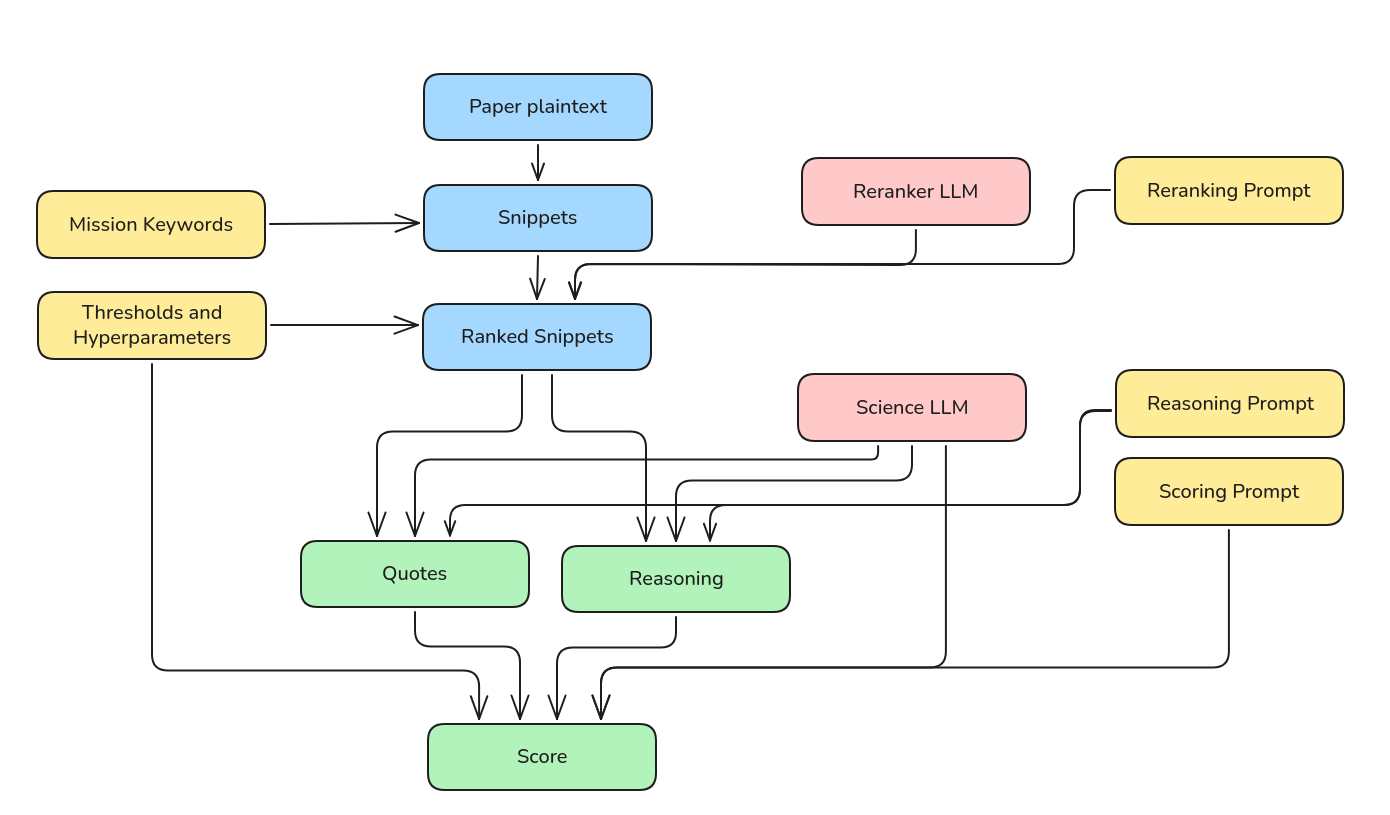}
    \caption{Schematic showing the system design for \amc{}. Note that the version of \amc{} adapted for TRACS does not separate the LLM generation of quotes, reasoning, and predictions; rather they are all output together. See the text in Section~\ref{sec:amc} and Appendix~\ref{sec:jwst-preprints} for a full discussion of differences.}
    \label{fig:schematic}
\end{figure*}

Figure~\ref{fig:schematic} shows a high-level overview of the \texttt{amc} system. 
The system classifies a single paper and a single telescope at a time.

First, \amc{} performs a keyword search to filter all mentions of telescope-related keywords, and we include surrounding context ($\pm 3$ sentences). This step effectively converts the body into a list of telescope-specific text snippets (Section~\ref{ssec:keywords}). Text snippets are then ranked by their relevance to the core question of ``is this a \{telescope\} science paper?'', and we only keep the most relevant snippets (Section~\ref{ssec:reranking}). These top-ranked snippets are subsequently passed to an LLM, which is prompted to classify the paper types and provide quotes and supporting reasoning for its predictions (Section~\ref{ssec:prompts}). 
The specialized code used for TRACS is forked from \amc{} and can be found at \url{https://github.com/jwuphysics/tracs_wasp2025}.

Finally, we note that our LLM system design is strongly influenced by a prior task: classifying whether arXiv paper preprints contained JWST science. In Appendix~\ref{sec:jwst-preprints}, we describe how these earlier motivations shaped (and biased) the design of the \amc{}. Additional discussion of the limitations of \amc{} are discussed in Section~\ref{ssec:limitations}.

\subsection{Keyword Filtering on Full Text} \label{ssec:keywords}

We concatenate the title, abstract, and body as a single text input. 
We extract only the most relevant portions of the text by searching for keywords. First, we divide text into sentences by using the Punkt sentence tokenizer \citep{punkt} in the NLTK package \citep{nltk}. We then use a simple Python case-insensitive string search to identify sentences with keywords for the relevant telescope. We expand snippets to include the $n=3$ prior and following sentences (i.e., such that each snippet contains $2n+1$ sentences). If no keywords are found, then we automatically classify the paper as \texttt{not\_telescope}.

We note that our keywords prioritize high recall at the expense of low precision; in other words, we value keyword completeness to make sure that no important keywords are missed. However, this means that false positives are expected. For example, our simple string matching over ``COS'' (an instrument on the Hubble Space Telescope) will also trigger matches on the words ``cosmic'' or ``cosine.'' Therefore, it is essential that we guard against false positives by ranking text snippets according to their relevance.

\subsection{Reranking Excerpts} \label{ssec:reranking}

Rerankers are typically LLMs that determine the relevance of some text snippet for answering a specific question. In information retrieval systems or retrieval-augmented generation (RAG), a first-stage algorithm usually produces an initial ranking or filtering over relevant documents/snippets (e.g. via semantic similarity in an embedding space). Rerankers provide a second-stage ranking between the query and a smaller set of snippets; recent works have demonstrated them valuable for LLM systems in astronomy and science more broadly \citep{Pathfinder,chen2025scirerankbenchbenchmarkingrerankersscientific,2025arXiv250707155X}. 

We implement a custom reranker solution\footnote{We were unaware (until the time of writing) that this reranking approach had been proposed in the literature before \citep[see e.g.][]{liang2023holistic}.} that achieves similar performance to leading commercial products (e.g., Cohere Reranker v3.5; based on a few informal assessments).
We use a lightweight, general-purpose, non-thinking model (\texttt{gpt-4.1-nano}) with a restricted vocabulary (``Yes'' and ``No'') that outputs logits (log-probabilities) between 0 and 1. 
Using a short reranker prompt, we task this model to identify whether each individual snippet discusses the telescope in a way that may be used to classify the paper type.
One of the main goals of this step is to remove accidental and unrelated keyword matches.

Once every snippet has a reranker score, we can sort them and/or filter out irrelevant snippets.
We keep up to $k=15$ top-ranked snippets in order to reduce the amount of text that is sent to the next LLM call.

\subsection{LLM Classifications} \label{ssec:prompts}

We combine the filtered text snippets together along with their reranker scores. The scores can serve as another reference for whether snippets are useful for determining the paper type.

We use \texttt{gpt-5-mini} with minimal reasoning effort to make the final classification as a structured output. The LLM prompt contains the top-ranked snippets and their scores, and it defines the different paper types and provides some examples. In addition to predicting boolean classes for the \texttt{science}, \texttt{instrumentation}, \texttt{mention}, and \texttt{not\_telescope} paper types, the LLM is also prompted to supply the most relevant quotes and justify its reasoning. All structured outputs and their data types are constrained via a \texttt{pydantic} model schema (e.g., boolean predictions, a list of strings for the quotes, and a single text string for the reasoning).

In the \amc{} package, the quotes and justification are provided first, followed by a separate LLM call to predict the final score on these lines of thinking (see, e.g., Figure~\ref{fig:schematic}). However, because TRACS requires multiple classifications, we simplify the system so that all predictions and quotes/reasoning are output at the same time.
The original \amc{} also supports floating point values between 0 and 1 for scoring \texttt{science} paper types, which allows another hyperparameter to control the threshold for scoring science papers. For TRACS, we simplified the system by using boolean values for each prediction.




\section{Results} \label{sec:results}

We briefly present some limited results on the TRACS test set. Our best score in terms of $F_1$ is 0.84 on the held-out test set, enough for a third-place rank according to the Kaggle leaderboard. In Appendix~\ref{sec:examples}, we show the \amc{} JSON-formatted outputs, including paper type examples for \texttt{science} (Listing~\ref{lst:science}), \texttt{instrumentation} (Listing~\ref{lst:instrumentation}), \texttt{mention} (Listing~\ref{lst:mention}), and \texttt{not\_telescope} (Listing~\ref{lst:not-telescope}). Based on a cursory review, these outputs seem accurate, the quotes do not suffer from hallucinations (although the risk is still present), and the provided reasoning largely appears to be faithful to its classification.

\subsection{Evaluating \amc{}}

In order to understand our system's strengths and weaknesses, we select $N = 100$ random entries from the training set, comprising 25 rows per telescope. This small, non-representative evaluation set enables us to investigate why our LLM system tended to make incorrect predictions. This random set is also able to surface potential issues with the dataset (see Section~\ref{ssec:errors}).  

\begin{figure*}
    \centering
    \includegraphics[width=\linewidth]{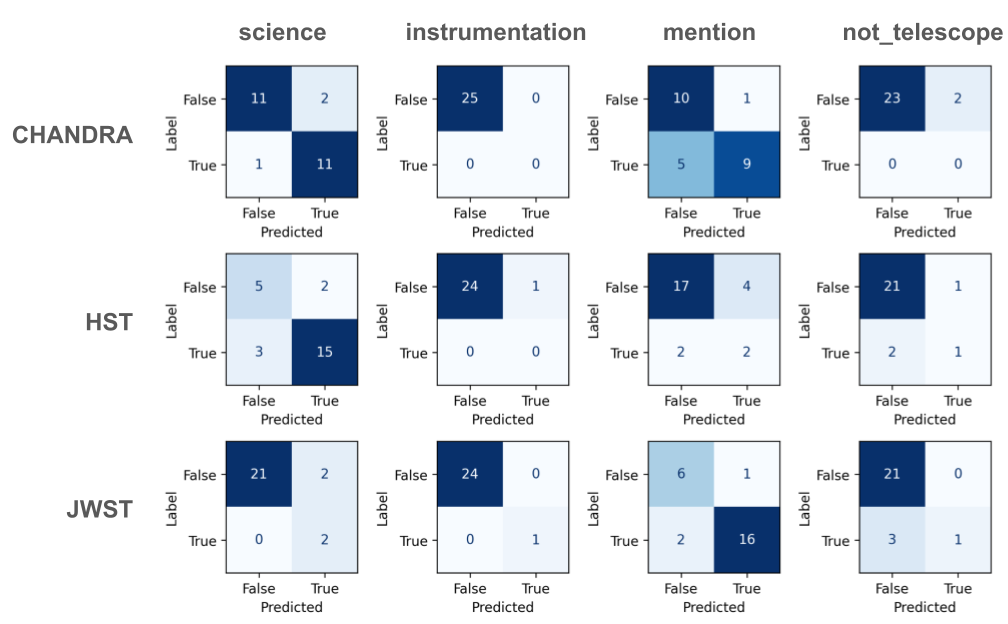}
    \caption{Binary confusion matrices, over a randomized subsample of 25 TRACS training set examples for each telescope (\textit{rows}), shown for each paper type (\textit{columns}) as a one-vs-all classification. Each confusion matrix shows true negatives (\textit{top left}), false positives (top right), false negatives (\textit{bottom left}), and true positives (bottom right).}
    \label{fig:confusion-matrices}
\end{figure*}

In Figure~\ref{fig:confusion-matrices}, we show confusion matrices displaying \amc{} predictions on the limited validation set, for all telescopes except \texttt{NONE}. Each column shows a paper type (denoted ``True'') against all other paper types (denoted ``False''). We note that some combinations of missions and paper types tend to succeed (e.g., \texttt{CHANDRA}/\texttt{science}) or fail more frequently (e.g., \texttt{CHANDRA}/\texttt{mention}). These confusion matrices are based on the same version of \amc{} as the final TRACS submission. 
However, we caution against overinterpreting results on this relatively small evaluation set.

\subsection{Performance on TRACS}

Our first submission to TRACS achieved a macro $F_1$ score of 0.80. At the time, the system included a few suboptimal settings, e.g., slightly misspecified prompts, or a non-zero reranker threshold which caused weak \texttt{mention} classes to occasionally be mislabeled (since the threshold might cause all text snippets to be filtered out, rendering a default verdict of \texttt{not\_telescope}).

After removing the reranker threshold and updating the prompts, we saw a modest increase in macro $F_1$ score to 0.84. We examined two of our higher-scoring sets of predictions, and used an LLM as a judge (\texttt{gpt-5-mini}) to resolve discrepancies between them and to issue final predictions; the performance remained at $F_1 = 0.84$.

The final LLM system took less than $24$ hours in wall-clock time to run, and incurred roughly $\$10$ in OpenAI costs. About $22\%$ of the cost is for reranking snippets with \texttt{gpt-4.1-nano}, $37\%$ is for processing top-ranked inputs with \texttt{gpt-5-mini}, and $41\%$ is for generating outputs with \texttt{gpt-5-mini}). Batch processing could lower some costs, but would necessitate an asynchronous pipeline, where we first perform all reranker calls, followed by all LLM classifications.


\subsection{Missing Data and Label Errors} \label{ssec:errors}

The dataset likely contains errors or uncertain classifications due to the imperfect nature of manually annotating bibliographic data, and the somewhat subjective nature of label distinctions. However, it is not possible to capture this uncertainty in the discrete classes. We also cannot measure the error rate directly, as there is no \textit{golden sample} against which we can compare. A golden sample would consist of papers that have been independently classified by multiple reviewers, where cases of disagreement are subject to deliberation and re-review until consensus is reached. Therefore, the error rate or uncertainty is unknown.

Through repeated evaluation, we can surface potential errors in the TRACS dataset.
While testing our LLM system on a small ($N=100$) subsample from the training data, we inspected all cases where the LLM prediction disagreed with the target label. Some of these appeared to be genuine error or ambiguity in the ground truth dataset, and we display them in Table~\ref{tab:errors}.

\begin{table*}[th]
    \centering
    \caption{Potential dataset issues found in a random selection of 100 labeled examples.}
    \label{tab:errors}
    \footnotesize
    \begin{tabular}{ll}
        \toprule
        \textbf{Bibcode} & \textbf{Notes} \\
        \midrule
        \texttt{2001AJ....122..598D} & Labeled as both \texttt{CHANDRA/science} and \texttt{CHANDRA/mention}. \\
        \midrule
        \texttt{2001ApJ...550..104Y}  & Missing body text, making it impossible to correctly classify as \texttt{HST/science}. \\
        \midrule
        \texttt{2002IJMPA..17.3446T} & Mentions ``Next Generation Space Telescope'' (the placeholder name for JWST), \\ & but the label is \texttt{JWST/not\_telescope} rather than \texttt{JWST/mention}.\\
        \midrule
        \texttt{2004RMxAC..20..215S} & Missing body text, making it impossible to correctly classify as \texttt{CHANDRA/science}. \\
        \midrule
        \texttt{2004fxra.book...89D} & Missing body text, making it impossible to correctly classify as \texttt{CHANDRA/science}. \\
        \bottomrule
    \end{tabular}
\end{table*}

We find that one paper is labeled as both \texttt{science} and \texttt{mention}, which (we assume) should not be possible. This classification may have resulted from human annotation error, or perhaps an accidental combination of a HST \texttt{mention} (as the paper is about the Hubble Deep Field) and CHANDRA \texttt{science}. Simultaneously conflicting labels like this can be easily filtered out by using boolean logic and some set rules.
We find another paper that mentions the ``Next Generation Space Telescope,'' the original name for JWST. Arguably, this paper should be considered a JWST mention, but is instead labeled as \texttt{not\_telescope}.

Three papers are missing their body text; they are only described by their titles, abstracts, and other metadata. For each of these three entries, we verify that (within the TRACS data) there is no mention or science/instrumentation of the candidate telescope presented. Missing full body text is often a symptom of complex publisher licensing agreements, and it may not always be possible to procure the full data. In any event, such entries do not contain sufficient data for making accurate predictions.

\subsection{Limitations of \amc{}} \label{ssec:limitations}

As noted above, the \amc{} system is designed to be general. Although we have specialized the code for the TRACS task, there are additional adjustments that could lead to improved performance. For example, the multiclass predictions would benefit from dedicated prompts for each paper type. The current system effectively uses the same prompts for each telescope, which might also limit its performance.

We also note that \amc{} is at the mercy of our keyword filtering. If we miss any telescope keywords, then it is possible to filter out relevant snippets, which could jeopardize the prediction task performance. Frequent keywords could be empirically learned using traditional NLP techniques like term frequency (TF; \citealt{term-freq}) normalized by its document frequency (i.e., TF-IDF; \citealt{tfidf}. The reranker step could potentially be replaced by a simple first-pass classifier using TF-IDF or another data-driven approach.

\section{Discussion} \label{sec:discussion}

LLMs are becoming pervasive throughout astronomy. 
Quantitative benchmarks \citep{2025arXiv251005016C,2025arXiv250520538J,2025A&C....5100893T} and human-centered studies \citep{2024arXiv240920252F,2024arXiv240520389W,hyk2025from} deliver complementary evaluations for how to successfully deploy LLMs for real-world benefit in astronomy. There is also rapid adoption of LLMs for navigating through and interacting with the astronomy literature \citep{2023RNAAS...7..193C,Pathfinder}, which is particularly salient for WASP/TRACS.

As researchers are in the midst of a fundamental shift of how they interact with literature, we discuss a future vision of how the astronomical community may leverage LLMs to augment or automate bibliographies (Section~\ref{ssec:vision}), how AI systems can assist in evaluating or improving our ground truth datasets (Section~\ref{ssec:better-ground-truths}), and how the \amc{} software we presented could be improved further in future work (Section~\ref{ssec:future}). 

\subsection{Scalable, AI-Supported Bibliographies} \label{ssec:vision}

We have shown that compiling telescope bibliographies can be assisted by or partially automated with LLMs.
LLM developments are built on traditional NLP techniques, which have already been vital for astronomical literature review \citep{Pathfinder} and detecting usage of telescopes/facilities \citep[e.g., using TF-IDF,][]{2025AJ....169...42A}.
While LLMs can be more expensive to put into production relative to simple NLP techniques or specialized fine-tuned models \citep[e.g., SciBERT,][]{SciBERT}, LLMs that have been pre-trained on trillions of tokens of general text are also capable of in-context learning via zero- or few-shot demonstrations \citep{gpt2,gpt3}. 
Modern LLMs also have longer context windows, enabling them to ingest multiple text snippets (or even entire documents at a time). This feature is particularly valuable if the telescope classification depends on nuanced text snippets buried within the body (i.e., often the case for \textit{archival} data sets, and rare for \textit{flagship} NASA missions; see Appendix~\ref{sec:other-bib}).

AI systems can still be extremely useful even if manual vetting of bibliographies is necessary.
We have designed \amc{} to have high recall, so it can confidently remove from consideration papers that have no chance of being \texttt{mention} paper types. Accurate labels ($F_1 > 0.8$) can dramatically save human time and mental energy.

\subsection{Errors and Ground Truths} \label{ssec:better-ground-truths}

When creating LLM-augmented bibliographies at massive scale, it is imperative to understand how the LLM is susceptible to errors, and/or if those errors originate from the LLM or from the dataset. For TRACS, our analysis of a small subsample in Section~\ref{ssec:errors} resulted in direct performance gains; we exposed some issues with our system, as well as errors in the dataset. 

We emphasize the value in compiling a golden sample with consensus reviews, even if this dataset is much smaller compared to the archival set of (single-pass) human classifications. In our prior work (see Appendices~\ref{sec:jwst-preprints} and \ref{sec:other-bib}), we have relied on a golden sample with about $N\sim 100$ examples to serve as a benchmark for improving the LLM system (Shaw et al., \textit{in prep}). Crucially, it also serves as a measure of \textit{human performance}, which is often incorrectly assumed to be perfect. By setting human error rates as the error ``floor,'' we can quantify a goal for LLMs to achieve.

AI augmentation can also facilitate a better understanding of our datasets.
For example, LLMs can easily comb through a large number of negative classes from \textit{historical} datasets, and surface candidate missing papers or other errors (e.g., Section~\ref{ssec:errors}). 
An LLM can be vital for efficiently constructing such a golden sample dataset.

\subsection{Future Improvements} \label{ssec:future}

Our solution for the TRACS task can likely benefit from additional optimization. In particular, other LLMs can help iteratively optimize the prompts used to guide the (TRACS-specific) \amc{} code, by using meta-optimizers \citep[see, e.g.,][]{mipro,gepa} in a prompt compilation framework like DSPy \citep{dspy}. Given the large TRACS training data set, meta-optimization could be costly, and may be precariously sensitive to the training label quality. However, meta-optimization could also produce (as a byproduct) empirical definitions of paper types like \texttt{science} or \texttt{instrumentation}, which could be valuable for comparing against explicit definitions that bibliographers have historically adopted. 

Another option is to use AI agents: LLMs that can call tools in a loop in order to accomplish a task.\footnote{For one definition of an AI ``agent'' that we like, see \url{https://simonwillison.net/2025/Sep/18/agents/}} Even though an AI agent might access the same tools that we have described in Section~\ref{sec:amc}, e.g., keyword search, reranking, filtering, or summarization, the LLM's \textit{agency} means that it can decide when and how to use such tool calls. The LLM agent can also maintain a working memory, allowing it to determine whether it has enough information to make a classification; for instance, if it finds immediate evidence that the paper presents scientific results, then the agent can stop the analysis and classify the paper as \texttt{science}.

Finally, we may wish to deploy \textit{smaller}, specialized models for this task because they can be run locally and perhaps at lower costs. For example, our keyword filtering and reranking steps are somewhat reminiscent of ``late-interaction'' retrieval mechanisms \citep[e.g., ColBERT,][]{colbert}, and it may be advantageous to substitute those steps with more lightweight model like ColBERT. We might simplify further by substituting this initial stage with classical NLP algorithms like TF-IDF. Models with specialized tokenizers for scientific literature like SciBERT \citep{SciBERT} may also prove to be beneficial for parsing the astronomical literature.

\section{Summary} \label{sec:summary}

We have presented \amc{}, an LLM-based system that can automatically categorize real astronomical papers into specific labels. 
Using a specialized instance of \amc{}, we demonstrate strong performance ($F_1 = 0.84$) and secure third place on the TRACS shared task \citep{grezes-etal-2025-tracs}. 
Our tool is also valuable for evaluating labeled data quality, as it provides reasoning and supporting quotes to justify its predicted labels.
Given the growing volume of papers, as well as the rising capabilities of LLMs, we believe that AI tools represent scalable solutions for accomplishing or assisting with this task. 

In the future, however, LLMs may completely obviate the need for predefined ``classifications'' that comprise current paper types; instead, we may be able to \textit{directly} ask LLMs questions like: ``How many papers present ground-based follow-up observations for targets initially discovered with HST?'' or ``How did the fraction of Chandra \textit{archival} science papers change between 2010 through 2025?'' We envision that, by exploiting the capabilities of AI systems, library scientists can study a broader range of bibliographic questions than ever before.

\section*{Ethical Disclosure}
All of this text was written solely by the authors. The document was partially reviewed by LLMs, primarily \texttt{gpt-5} and \texttt{Gemini 2.5 Pro}, in order to surface issues in clarity and prose. Some of the code in the associated repository is generated by LLMs, primarily via Claude Code. Having validated the software and results, the authors take full responsibility and ownership over the results presented here.

\section*{Acknowledgments}
We acknowledge helpful conversations with STScI staff, including Brian Cherinka, Dick Shaw, and Jinmi Yoon.
Some of the ideas about combining the LLM system with TF-IDF and SciBERT are based on exchanges with Madhusudhana Naidu after the end of the TRACS competition.

\bibliography{main}

\onecolumn
\appendix

\section{Examples of \amc{} Structued Outputs for the TRACS Test Set} \label{sec:examples}

In Listings \ref{lst:science}, \ref{lst:instrumentation}, \ref{lst:mention}, and \ref{lst:not-telescope}, we show several representative examples of \amc{} outputs for the TRACS test dataset.

\begin{listing}[ht]
\begin{lstlisting}[language=json, captionpos=b]
"2024ApJ...977..173C_JWST": {
  "id": "2024ApJ...977..173C_JWST",
  "bibcode": "2024ApJ...977..173C",
  "telescope": "JWST",
  "classification": {
    "telescope": "JWST",
    "science": true,
    "instrumentation": false,
    "mention": false,
    "not_telescope": false,
    "quotes": [
      "we present JWST MIRI observations of the hydrocarbon-rich source, DoAr 33, a 1.1 M star.",
      "We detect the emission of H 2 O, CO 2 , OH, and C 2 H 2 , including its isotopologue 13 C 12 CH 2 , C 4 H 2 , and tentatively CH 4 and HC 3 N, for all of which we retrieve co
lumn densities, temperatures, and emitting areas, as described in Section 2 .",
      "We detect the presence of H 2 O, CO 2 , OH, C 2 H 2 , HCN, C 4 H 2 , and tentatively CH 4 and HC 3 N in the JWST MIRI-MRS spectrum of the solar-mass star DoAr 33.",
      "JWST/MIRI Detection of a Carbon-rich Chemistry in the Disk of a Solar Nebula Analog",
      "Figure 1. JWST MIRI-MRS spectrum of DoAr 33."
    ],
    "reasoning": "The excerpts clearly describe original JWST MIRI-MRS observations of the target DoAr 33 and report new detections and quantitative retrievals (column densities, te
mperatures, emitting areas) derived in this paper. These are new scientific results based on JWST data, so 'science' is True. The text discusses data reduction using a custom JDISCS p
ipeline and calibration references but does not present new instrument design, calibration method development as the primary focus or a technical/instrumentation paper, so 'instrument
ation' is False. Because the paper actively uses JWST data to produce new results, it is not merely a mention; therefore 'mention' is False. There is no indication that references to
JWST are false positives, so 'not_telescope' is False."
  }
}
\end{lstlisting}
\caption{\amc{} output for an \texttt{science} paper type. \label{lst:science}}
\end{listing}

\begin{listing}[ht]
\begin{lstlisting}[language=json]
"1996SPIE.2808..650F_CHANDRA": {
  "id": "1996SPIE.2808..650F_CHANDRA",
  "bibcode": "1996SPIE.2808..650F",
  "telescope": "CHANDRA",
  "classification": {
    "telescope": "CHANDRA",
    "science": false,
    "instrumentation": true,
    "mention": false,
    "not_telescope": false,
    "quotes": [
      "Modeling the diffraction efficiencies of the AXAF high-energy transmission gratings: II",
      "In order to characterize the instrumentation on AXAF, each of the science instrument teams carries out sub-assembly calibrations.",
      "For the high energy transmission grating (HETG) group, this means individual measurements of the diffraction efficiencies of each of the 336 grating elements that goes into t
he completed HETG assembly."
    ],
    "reasoning": "The excerpt discusses calibration and modeling of the AXAF high-energy transmission gratings (HETG), measurements, data reduction and instrumentation characterization. AXAF is the original name for the Chandra X-ray Observatory, and the content is clearly about instrument performance and calibration rather than presenting new astrophysical results. Therefore instrumentation = True. It does not present new scientific/astronomical analyses using telescope observations, so science = False. Because the paper actually discusses the telescope/instrument technical work, it should not be classified as a mere mention or as not_telescope."
  }
}
\end{lstlisting}
\caption{\amc{} output for an \texttt{instrumentation} paper type. \label{lst:instrumentation}}
\end{listing}

\begin{listing}[ht]
\begin{lstlisting}[language=json]
"2024A&A...691A.352W_JWST": {
  "id": "2024A&A...691A.352W_JWST",
  "bibcode": "2024A&A...691A.352W",
  "telescope": "JWST",
  "classification": {
    "telescope": "JWST",
    "science": false,
    "instrumentation": false,
    "mention": true,
    "not_telescope": false,
    "quotes": [
      "an unbiased mid-infrared survey observing H-dominated metal-enriched white dwarfs making use of the much improved sensitivity of JWST, compared to previous facilities, would likely reveal the true population of discs.",
      "infrared spectroscopy can identify the mineralogy of discs, with silicates having been detected using Spitzer and JWST ( Jura et al. 2007 ; Swan et al. 2024 )."
    ],
    "reasoning": "The excerpt discusses JWST in the context of future or potential surveys and improved sensitivity compared to previous facilities, and cites JWST detections (Swan et al. 2024) as literature. There is no indication that this paper presents new JWST observational data or original data analysis; it instead suggests that an unbiased survey with JWST would be useful and refers to prior JWST detections. This fits the 'mention' category (discussion of future/planned observations and citing others). It does not describe instrumentation or calibration, and the references to JWST clearly refer to the telescope, so not_telescope is false."
  }
}
\end{lstlisting}
\caption{\amc{} output for a \texttt{mention} paper type. \label{lst:mention}}
\end{listing}

\begin{listing}[ht]
\begin{lstlisting}[language=json]
"2008ApJ...673..657M_HST": {
  "id": "2008ApJ...673..657M_HST",
  "bibcode": "2008ApJ...673..657M",
  "telescope": "HST",
  "classification": {
    "telescope": "HST",
    "science": false,
    "instrumentation": false,
    "mention": false,
    "not_telescope": true,
    "quotes": [
      "Hubble diagram showing the magnitude deviation relative to an empty universe, for the three models considered.",
      "Fig. 4. - Hubble diagram showing the magnitude deviation relative to an empty universe, for the three models considered. The dotted lines separate the Tonry et al. ( 2003 ) data ( left ) from the mock data ( right )."
    ],
    "reasoning": "The excerpts discuss a 'Hubble diagram' and cosmological supernova analyses; there is no mention of the Hubble Space Telescope or HST observations, instruments, calibrations, or use of HST data. The term 'Hubble' here refers to the Hubble diagram (relation to Hubble expansion), which is a false positive for the telescope name. Therefore this is not about the HST telescope (not_telescope = True). Because it does not discuss any telescope data usage or instrumentation, science and instrumentation are False. It also is not merely a mention of the telescope, so mention = False. The provided exact substrings showing 'Hubble diagram' are included as supporting quotes."
  }
}
\end{lstlisting}
\caption{\amc{} output for a \texttt{not\_telescope} paper type. \label{lst:not-telescope}}
\end{listing}

\section{The JWST Preprints Automation} \label{sec:jwst-preprints}

The design of \amc{} was initially devised for a specific task: automatically checking whether new arXiv preprints feature JWST science.
While identifying JWST science preprints is useful on its own, it is also necessary for another goal: determining whether JWST science publications are accompanied by digital object identifiers (DOIs), which offers downstream benefits to individual scientists \citep{2012ASPC..461..763H} and the broader science community \citep{10.1371/journal.pone.0104798}. Therefore, we built an earlier version of \amc{} for automatically classifying whether a JWST preprint is \texttt{science} or not.\footnote{\url{https://github.com/spacetelescope/jwst-preprint-automator}}
We achieved $\bm{F_1 \approx 0.95}$ on real-world tests, using a held-out golden sample dataset with $N=114$ papers, and this system is currently being used in production at STScI.

There are two major differences between the JWST preprints automation and the TRACS task: (1) the former downloads PDF files from arXiv and converts them into a single plain-text body, while the latter provides more cleanly separated metadata and body text (when available); and (2) the former task is only concerned with identifying \texttt{science} papers, whereas the latter solicits binary classifications for \texttt{science}, \texttt{instrumentation}, \texttt{mention}, and \texttt{not\_telescope}.

There are several important implications due to distinction \#1. For example, we do not have a programmatic method for separating abstracts or other titles from the body, and so we must always proceed with the full body text. 
Another consequence is that the references section is included in the plain-text body extract from arXiv preprints. Thus, the body text contains references to titles of \textit{other} papers, which can sometimes mimic sentences that appear to support a JWST science classification. 

Because we focus only on classifying whether a JWST paper is \texttt{science} (distinction \#2), we break down the LLM output into two stages (see Figure~\ref{fig:schematic}. First, we write out a specialized prompt with in-context examples of low and high \texttt{science} scores, and prompt the LLM to output reasoning and supporting quotes. Then, given its provided justification and quotes, we force the LLM to output a \texttt{science} score as a floating point number between 0 and 1.

\section{Other Mission Bibliographies} \label{sec:other-bib}
We note that NASA flagship missions tend to be mentioned in the paper abstract, or even in the title. Thus, for TRACS and for JWST preprints automation, it is often possible to accurately classify papers using just the abstract, a tiny fraction of the available text per entry. 

However, we designed \amc{} based on our experience performing bibliometrics for other missions, including TESS, Pan-STARRS, and GALEX, which all enable \textit{archival} science well after the telescope ceases operations \citep[e.g.,][]{2019BAAS...51g.105P}. However, these telescopes---and, in general, archival science usage of telescopes---are less likely to appear in the title or abstract. Therefore, parsing the much longer body text is imperative for correctly predicting those telescopes' paper types. We design \amc{} so that it can be used just as easily to compute bibliometrics for non-flagship telescopes and/or archival science.

\end{document}